\documentclass[aps,pra,onecolumn,12pt,showpacs]{revtex4-1}
\usepackage[utf8x]{inputenc}
\usepackage[T1]{fontenc}
\usepackage[english]{babel}
\usepackage{eurosym}
\usepackage{amsmath, amsthm, amscd, amssymb,amsbsy,amsfonts,wasysym}
\usepackage[dvips]{graphicx}
\usepackage{fancyhdr}
\usepackage{vmargin}
\usepackage{multirow,framed}
\usepackage[table]{xcolor}
\usepackage{makeidx}
\usepackage{color}

\begin{document}

\title{Progress towards an effective non-Markovian description of a system interacting with a bath}   
\author{L. Ferialdi}
\email{ferialdi@math.lmu.de}
\affiliation{Mathematisches Institut, Ludwig-Maximilians-Universit\"at, Theresienstr. 39, 80333 Munich.}
\author{D. D\"urr}
\email{duerr@math.lmu.de}
\affiliation{Mathematisches Institut, Ludwig-Maximilians-Universit\"at, Theresienstr. 39, 80333 Munich.}
\begin{abstract}
We analyze a system coupled to a bath of independent harmonic oscillators. We transform the bath in chain structure by solving an inverse eigenvalue problem. We solve the equations of motion for the collective variables defined by this transformation, and we derive the exact dynamics for an harmonic oscillator in terms of the microscopic motion of the environmental modes. We compare this approach to the well-known Generalized Langevin Equation and we show that our dynamics satisfies this equation.
\end{abstract}
\maketitle

\section{Introduction}
Ultra-fast physical processes captured the attention of the scientific community because they arise in different physical situations ranging from chemistry~\cite{Pometal} to condensed matter~\cite{Xiaetal}, to bio physics~\cite{Thoetal}. From the theoretical point of view these processes are described by open systems, i.e. systems interacting with the surrounding environment. Most commonly, for an effective description open quantum systems are approximated by Markovian dynamics, which require a large separation of time-scales between the system and the environment~\cite{BrePet02}. Ultra-fast processes are those for which the time-scale of the relevant system is about of the same order as that of the bath into which it is immersed. Therefore an effective characterization of such processes - if existing at all - must rely on other modes of description, which most likely need to be non-Markovian. This justifies the growing interest in non-Markovian open quantum system dynamics. It is important to stress that non-Markovian descriptions can only be effective on short time-scales: this feature will play a crucial role in choosing the best strategy to tackle non-Markovian dynamics.

The model most widely used to describe open quantum systems is the \lq\lq independent oscillators\rq\rq~(IO) model~\cite{Zwa73,CalLeg83,ForLewOco88a}. In this model the system is bilinearly coupled to a bath of independent harmonic oscillators.  This model has been thoroughly studied, and it proved fundamental for the description of Markovian and non-Markovian quantum Brownian motion~\cite{CalLeg83,GraTal83,HuPazZha92}. Considering the Heisenberg equations of motion of this model, one can derive a Generalized Langevin Equation (GLE), that gives a phenomenological description of how the environment affects the system~\cite{Mor65,Kub66,Zwa73,BenKac81,ForKac87,ForLewOco88b}. The structure of the GLE is perfect for deriving the Markovian limit, which in the classical regime recovers the Langevin equation for Brownian motion. In short, the GLE is suitable to analyze thermal effects, diffusive effects, and the fluctuation-dissipation relations due to the environment. However, as we shall argue the GLE is not suited in situations when the timescale separation is small and non-Markovian features are dominant.
 Indeed, the picture given by the IO model and the GLE only tells us that the interaction with the environment gives rise to some non-Markovian/memory effects, but it is not clear how one can capture them in an effective way.
 
In order to tackle this problem, we will consider another representation which suits better the analysis of short time dynamics. As we shall show the IO model indeed is not suitable for this scope, because the influence of the interaction between the system and the environment is not \lq\lq time\rq\rq-ordered as all the environmental oscillators act \lq\lq at the same time\rq\rq~on the system. A better representation from this point of view is a chain model, i.e. a system interacting with a chain of first neighbor interacting harmonic oscillators~\cite{Rub60,ForKacMaz65}. 
This model allows to order the influence of the interaction in time, and one can have a clearer physical picture of what is going on: the first oscillator of the chain will first affect the system, and only at a later time the influence of the second oscillator of the chain will reach the system, and so on and so forth.
As we will show, such a model allows to express the system dynamics in terms of the microscopic motion of the bath constituents [cf. Eq.~\eqref{xsol}].
As IO models, due to their structural closeness to GLE are mostly considered in applications, we ask the question: How can we achieve a chain structure from an underlying model of independent oscillators?
The neatest way to perform the transformation between these two descriptions is by solving an Inverse Eigenvalue Problem (IEP). These problems are well known in the field of vibration theory, and the literature is vast (see e.g.~\cite{Gla04,ChuGol05} and references therein).

Our analysis applies both to Classical Mechanics (if one consider phase-space variables) and to Quantum Mechanics (if one considers Heisenberg equations of motion). 
The paper is organized as follows: in Sec. 2 we introduce the two models for the environment, and we solve the IEP that univocally determines the parameters of the chain. In Section 3 we solve the equations of motion of the chain model in terms of the environmental modes, obtaining the exact dynamics of the system. We eventually compare our results with the phenomenological GLE.


\section{Independent oscillators and chain models}
 We consider an open quantum system made of a particle bilinearly interacting with an environment of $N$ independent harmonic oscillators. We assume the environment to be arbitrarily big but finite. The Hamiltonian describing this \lq\lq independent oscillators\rq\rq~(IO) model reads:
\begin{equation}\label{Hio}
H_{\text{\tiny{IO}}}=\frac{p^2}{2M}+V(x)+x\sum_{k=1}^N c_k q_k+\sum_{k=1}^N \frac{1}{2}\left(p_k^2+\omega_k^2q_k^2\right)\,,
\end{equation}
where $x,p$ are the position and momentum operators of the relevant system, $c_k$ are positive constants, and $q_k, p_k$ are position and momentum operators of the environmental oscillators with proper frequency $\omega_k$. $V(x)$ is a generic renormalized potential, which includes the term $x^2\sum_k c^2_k/2\omega_k^2$ that guarantees boundedness and translation invariance of the model. Note that this implies that also an initially free particle will always acquire an oscillatory behavior due to the interaction with the environment. 
Introducing the vector of the position operators of the environmental oscillators ${\mathbf q^{T}}=(q_1,\ldots q_N)$, one can write their Heisenberg equations of motion as follows:
\begin{equation}\label{motio}
\frac{d^2}{dt^2}{\bf q}(t)=-\boldsymbol{\omega}\cdot{\bf q}(t)\,,
\end{equation}
where $\boldsymbol{\omega}$ is the diagonal matrix of the oscillators frequencies: $\boldsymbol{\omega}=\mathrm{diag}(\omega_1^2,\ldots\omega_N^2)$, with $\omega_1<\omega_2<\dots\omega_N$.
Substituting the system~\eqref{motio} in the Heisenberg equation for $x$, one obtains the following GLE~\cite{Zwa73,CalLeg83}:
\begin{equation}\label{GLE}
\ddot{x}(t)+\int_0^tds\,\eta(t-s)\dot{x}(s)+V'(x)=g(t)\,,
\end{equation}
where the prime denotes differentiation with respect to $x$. The kernel $\eta(t-s)$ is called friction kernel, since in the Markov limit the integral term becomes the friction constant of the Langevin equation. The stochastic force $g(t)$ depends on the initial conditions of the bath operators and is responsible for the diffusive behavior. The explicit but cumbersome expressions for $\eta(t-s)$ and $g(t)$ can be easily derived in terms of the IO parameters~\cite{CalLeg83,ForLewOco88b}. Equation~\eqref{GLE} can be taken as the starting point for Markovian and non-Markovian descriptions. The Markovian description would arise when there is little memory, i.e. $\eta(t)$ is close to a Dirac delta and $g(t)$ is close to Brownian motion. Non-Markovian dynamics arise of course for kernel not approximative delta-like.

We now focus on the alternative mode of description - a chain model for the bath - which turns out to be more suitable to study the short time behavior of the dynamics. A system interacting with a chain of first-neighbor interacting harmonic oscillators is described by the following Hamiltonian:
\begin{equation}\label{Hchain}
H_{\text{\tiny{CHAIN}}}=\frac{p^2}{2M}+V(x)+DxX_1+\sum_{k=2}^N D_{k-1} X_{k-1}X_k+\sum_{k=1}^N \frac{1}{2}\left(P_k^2+\Omega_k^2X_k^2\right)\,.
\end{equation}
Here $x$ is the position of the system, a tracer particle, and $D_k$ are positive coupling constants, $X_k$ are the position operators of the chain oscillators, $P_k$ are their conjugated momenta, and $\Omega_k$ their frequencies.
The Heisenberg equations for $X_k$ read
\begin{equation}\label{motchain}
\frac{d^2}{dt^2}{\bf X}(t)=-\boldsymbol{T}\cdot{\bf X}(t)\,,
\end{equation}
where ${\mathbf X^{T}}=(X_1,\ldots X_N)$, and ${\mathbf T}$ is the following tridiagonal matrix:
\begin{equation}\label{tri}
{\mathbf T}=\left(
\begin{array}{cccc}
\Omega_1^2& -D_1& 0&\dots\\
-D_1&\Omega_2^2 &-D_2&\dots\\
0&-D_2& \Omega_3^2&\dots\\
\vdots &\vdots &\vdots & \ddots
\end{array}\right)\,.
\end{equation}
We may think of~\eqref{Hchain} as defining a microscopic model in its own rights, but in this paper we wish to take as fundamental the IO model and aim at achieving a description in terms of~\eqref{Hchain}. We shall refer to the chain description as if it were a model and call the $X_k$ the environment. Accordingly, we want to be able to relate our subsequent analysis of the chain model to the IO model at any time. In order to do so, we require the Hamiltonians $H_{\text{\tiny{IO}}}$, $H_{\text{\tiny{CHAIN}}}$ to be \lq\lq equivalent\rq\rq, in the sense that they give the same dynamics for the tracer particle $x$. Accordingly, the parameters entering $H_{\text{\tiny{CHAIN}}}$ are not free, but they have to be particular functions of the parameters of $H_{\text{\tiny{IO}}}$. Moreover, the chain oscillators $X$ have to be specific linear combinations of the independent oscillators $q$. Let us introduce an orthogonal $N\times N$ matrix $\mathbf{O}$, and define
\begin{equation}\label{X}
X_j= \sum_k O_{jk}q_k\,.
\end{equation}
In order to satisfy our requirements, the matrix $\mathbf{O}$ has to be such that the equations~\eqref{motio} and~\eqref{motchain} are equivalent. Substituting Eq.~\eqref{X} in~\eqref{motchain} one easily finds that this requirement reduces to that of determining the orthogonal matrix $\mathbf{O}$ such that:
\begin{equation}\label{IEP}
\mathbf{T}=\mathbf{O}\cdot\boldsymbol{\omega}\mathbf{O}^{T}\,.
\end{equation}
The problem of determining a matrix starting from its eigenvalues is known under the name of Inverse Eigenvalue Problem (IEP).
Kindred relations between the two environments have already been considered in the literature. In~\cite{Hugetal09} the authors define  \lq\lq collective modes\rq\rq~through a transformation that makes use of a hierarchical baths construction. The parameters of the chain are not obtained analytically, but fitted with some experimental data. A similar hierarchical transformation for an infinite chain is used in~\cite{Maretal11}, where the authors derive (formally) the chain parameters from the propagator of the initial GLE. Another approach for infinite chains has been proposed in~\cite{Chietal10}, where the collective modes are obtained exploiting the properties of orthogonal polynomials.

\subsection{Inverse Eigenvalue Problem}
IEPs arise in different areas of theoretical and applied sciences , like e.g. vibration theory, control theory, particle physics, geophysics, and engineering~\cite{Gla04,ChuGol05}. In general, an IEP consists in finding the entries of a matrix ${\mathbf T}$, starting from its eigenvalues and some additional initial data. In our case the matrix ${\mathbf T}$ is a Jacobi matrix, i.e. positive semi-definite, symmetric, tridiagonal matrix. A $N \times N$ Jacobi matrix has $2N-1$ free entries, while the IO model gives us $N$ conditions (the eigenvalues $\omega_k$). In order to find a unique solution to our IEP we need $N-1$ supplementary conditions, which are obtained by exploiting the equivalence between $H_{\text{\tiny{IO}}}$ and $H_{\text{\tiny{CHAIN}}}$. In particular, matching the interaction terms of Eqs.~\eqref{Hio} and~\eqref{Hchain} one finds that $DX_1=\sum_kc_k q_k$, which by means of Eq.~\eqref{X} implies $O_{1k}= D^{-1}c_k$. The knowledge of the first line of ${\mathbf O}$ provides enough conditions to uniquely solve the Jacobi IEP. The algorithm to determine ${\mathbf O}$ and ${\mathbf T}$ is standard so we do not repeat it here. We do refer only to the results essential to this paper, further details can be found in~\cite{Gla04,ChuGol05}.
The entries of the matrix $\mathbf O$ read 
\begin{equation}\label{O}
O_{jk}=\left(\prod_{l=1}^{j-1}D^{-1}_{j-1}\right)P_{j-1}(\omega_k)\,,
\end{equation}
where $P_j(\lambda)$ is the characteristic polynomial of the $j$-th leading principal minor of $\mathbf T$, evaluated in $\lambda$. Note that the explicit expressions for the $P_j$ are determined recursively exploiting the following recurrence relation:
\begin{equation}\label{recP}
P_{j+1}(\lambda)= (\Omega_j^2-\lambda)P_j(\lambda)-D_j^2P_{j-1}(\lambda)
\end{equation}
with $P_{-1}=0$.
Once the transformation matrix ${\mathbf O}$ is determined, the entries of ${\mathbf T}$ are given by the following relations:
\begin{eqnarray}
\Omega_j^2&=&\sum_k\omega_k^2O_{jk}^2\\
D_j&=&-\sum_k\omega_k^2O_{jk}O_{j+1k}\,.
\end{eqnarray}
These equations complete the set of parameters that we will need in the following discussion. From now on we will consider the matrix ${\mathbf T}$ as known, i.e. as fully determined in terms of the parameters of the IO model. The dynamics given by Eq.~\eqref{Hchain} with the parameters here defined is equivalent to that given by Eq.~\eqref{Hio}. Moreover, note that the $\Omega_j$ are not necessarily ordered with respect to the index $j$.
 The chain oscillators $X$ manifestly represent a particular collective behavior of the independent oscillators. Unfortunately, their structure in terms of $q$ is rather cumbersome and does not allow for a straightforward physical interpretation. 

\section{Non-Markovian dynamics of an harmonic oscillator}
The aim of this section is to determine how the collective modes $X$ affect the dynamics of the system. For this we need to solve the set of equations of motion~\eqref{motchain}. In order to do so, we approximate the tracer particle potential $V(x)$ (cf.~\eqref{Hchain}) harmonically, i.e. we consider now an harmonic oscillator with proper frequency $\Omega$. The Heisenberg equations of the open system can then be explicitly written as follows:
\begin{eqnarray}\label{mot}
\frac{d^2}{dt^2}x(t)&=&-\Omega^2x(t)+D X_1(t)\\
\frac{d^2}{dt^2}X_{i}(t)&=&-\Omega_i^2X_{i}(t)+D_{i-1}X_{i-1}(t)+D_iX_{i+1}(t)\,,\quad 1\leq i\leq N-1\\
\frac{d^2}{dt^2}X_{N}(t)&=&-\Omega_N^2X_{N}(t)+D_{N-1}X_{N-1}(t)\,.
\end{eqnarray}
We rewrite these equations in integral form: 
\begin{eqnarray}\label{intmot}
x(t)&=&f_0+\int_0^t \frac{\sin[\Omega_0(t-s)]}{\Omega_0} DX_{1}(s)ds\\
\label{intmoti}X_i(t)&=& f_i(t)\!+\!\!\!\int_0^t\!\! \frac{\sin[\Omega_i(t-s)]}{\Omega_i} (D_{i-1} X_{i-1}(s)\!+\!D_i X_{i+1}(s))ds\,,\,\, 1\leq i\leq N\!-\!1\\
\label{intmotN}X_{N}(t)&=&f_N+\int_0^t \frac{\sin[\Omega_N(t-s)]}{\Omega_N} D_{N-1} X_{N-1}(s)ds\,,
\end{eqnarray}
where  we have relabeled $\Omega_0=\Omega$, $X_0=x$, and
\begin{equation}\label{init}
 f_i(t)=X_i(0)\cos[\Omega_it]+\dot{X}_i(0)\frac{\sin[\Omega_i t]}{\Omega_i}\,.
\end{equation}
In order to obtain the dynamics of $x(t)$ in terms of the $X_i$, we substitute recursively Eq.~\eqref{intmoti} in Eq.~\eqref{intmot}. One can prove that substituting the equations up to $X_n(t)$ (for every $n\leq N$), x(t) reads
\begin{eqnarray}\label{xn}
x(t)&=&\tilde{f}_n(t)+\sum_{i=1}^n\left(\prod_{l=0}^i\frac{D_l}{\Omega_l}\right)\frac{D_{i-1}}{D_i}\int_0^t K_{i}(t-s)X_{i-1}(s)ds\nonumber\\
&&+\left(\prod_{l=0}^n\frac{D_l}{\Omega_l}\right)\int_0^t K_{n}(t-s)X_{n+1}(s)ds\,,
\end{eqnarray}
where recursively
\begin{eqnarray}
\label{Ki} K_{i}(t-s)&=&\int_s^tK_{i-1}(t-l)\sin[\Omega_i(l-s)]dl\\
\label{tilf} \tilde{f}_i(t)&=&\tilde{f}_{i-1}(t)+\left(\prod_{l=0}^{i-1}\frac{D_l}{\Omega_l}\right)\int_0^t K_{i-1}(t-s)f_{i}(s)ds\,,\nonumber
\end{eqnarray}
with $K_0(t-l)=\sin[\Omega(t-l)]$, $\tilde{f}_0(t)=f_0(t)$, and $f_i(s)$ given by Eq.~\eqref{init}.
Equation~\eqref{xn} is obtained by induction. To see its correctness for $n=1$ substitute Eq.~\eqref{intmoti} for $X_1(t)$ in Eq.~\eqref{intmot} to find
\begin{eqnarray}
x(t)&=&f_{0}(t)+\frac{D}{\Omega}\int_0^t \sin[\Omega(t-s)]f_{1}(s)ds\nonumber\\
&&+\frac{D^2}{\Omega\Omega_1}\int_0^t K_{1}(t-s)x(s)ds+\frac{DD_1}{\Omega\Omega_1}\int_0^t K_{1}(t-s)X_{2}(s)ds\,.
\end{eqnarray}
This equation can be easily recast in form of Eq.~\eqref{xn}. Assume now that Eq.~\eqref{xn} is true for a generic $n\leq N$, and substitute Eq.~\eqref{intmoti} for $X_{n+1}(s)$ in the second line of Eq.~\eqref{xn}. After some simple manipulation one can show that Eq.~\eqref{xn} results for $n+1$, i.e.
\begin{eqnarray}\label{xn+1}
x(t)&=&\tilde{f}_{n+1}(t)+\sum_{i=1}^{n+1}\left(\prod_{l=0}^i\frac{D_l}{\Omega_l}\right)\frac{D_{i-1}}{D_i}\int_0^t K_{i}(t-s)X_{i-1}(s)ds\nonumber\\
&&+\left(\prod_{l=0}^{n+1}\frac{D_l}{\Omega_l}\right)\int_0^t K_{n+1}(t-s)X_{n+2}(s)ds\,.
\end{eqnarray}

It is important to note that equation~\eqref{xn} is exact. Although only $n$ equations have been substituted into $x(t)$, the second line of Eq.~\eqref{xn} contains all the information regarding the evolution of the remaining $N-n$ modes, encoded in $X_{n+1}(s)$. It is easy to show that, if one substitutes all the $N$ equations for  $X_i$, $x(t)$ satisfies the following equation:
\begin{equation}\label{xN}
x(t)=\tilde{f}_N(t)+\sum_{i=1}^N\left(\prod_{l=0}^i\frac{D_l}{\Omega_l}\right)\frac{D_{i-1}}{D_i}\int_0^t K_{i}(t-s)X_{i-1}(s)ds\,.
\end{equation}
Indeed, $D_N=0$ by definition. Accordingly, the second line of Eq.~\eqref{xn} is null.

This equation determines the dynamics of $x(t)$ in terms of the full set of $X_i$. One can identify two different contributions to the evolution of $x$: one is given by $\tilde{f}_{N}(t)$ that collects the initial conditions of the collective modes; the second is a purely non-Markovian integral contribution which involves the whole past evolution of the collective modes.

In order to study the evolution of $x(t)$, recall that $X_0=x$ and rewrite  Eq.~\eqref{xN} as follows:
\begin{equation}\label{xNint}
x(t)=\frac{D^2}{\Omega\Omega_1}\int_0^t K_{1}(t-s)x(s)ds+F_N(t)\,,
\end{equation}
where the function $F_N(t)$ collects all terms of Eq.~\eqref{xN} that do not depend on $x$: 
\begin{equation}\label{FN} 
F_N(t)=\tilde{f}_N(t)+\sum_{i=2}^N\left(\prod_{l=0}^i\frac{D_l}{\Omega_l}\right)\frac{D_{i-1}}{D_i}\int_0^t K_{i}(t-s)X_{i-1}(s)ds\,,
\end{equation}
Equation~\eqref{xNint} explicitly shows that the dynamics of $x$ is ruled by an integral equation. Furthermore, we stress that Eq.~\eqref{FN} can be rewritten in terms of the independent oscillators by exploiting Eq.~\eqref{X}:
\begin{equation}\label{Xs}
X_j(s)= \sum_k O_{jk}\left(q_k(0)\cos\omega_ks+\dot{q}_k(0)\frac{\sin\omega_ks}{\omega_k}\right)\,.
\end{equation}
Recalling the definition of Eq.~\eqref{Ki}, and performing the integration, one finds that the kernel $K_1(t-s)$ is
\begin{equation}\label{K1}
K_{1}(t-s)=\frac{\Omega_1\sin[\Omega(t-s)]-\Omega\sin[\Omega_1(t-s)]}{\Omega_1^2-\Omega^2}\,.
\end{equation}
Since this is a simple combination of two sine functions, Eq.~\eqref{xNint} can be solved using standard techniques~\cite{PolMan08}. The solution reads
\begin{equation}\label{xsol}
x(t)=F_N(t)+\frac{D^2}{\mu_1\mu_2(\mu_2^2-\mu_1^2)}\int_0^t \left(\mu_2\sin[\mu_1(t-s)]-\mu_1\sin[\mu_2(t-s)]\right)F_N(s)ds\,,
\end{equation}
where
\begin{equation}
\mu_{1,2}=\sqrt{\frac{1}{2}\left(\Omega^2+\Omega_1^2\pm\sqrt{\Delta}\right)}\,,\qquad\Delta=(\Omega^2-\Omega_1^2)^2-4D^2\,.
\end{equation}
Note that in order to have real values for $\mu_{1,2}$ and avoid multi-valued $x(t)$, one needs the condition $\Delta\geq 0$ to hold true. 
Equation~\eqref{xsol} is the main result of this paper: it displays the exact solution of our problem, representing the dynamics of an harmonic oscillator under the influence of a chain of $N$ harmonic oscillators. We stress that all the parameters entering this equation are known analytically. Furthermore, by means of Eqs.~\eqref{FN}-\eqref{Xs}, one can rewrite Eq.~\eqref{xsol} in terms of the $q_k$: this implies that for any sets of parameters and initial conditions of the IO model, Eq.~\eqref{xsol} also provides the exact dynamics for this model.
In this sense, Eq.~\eqref{xsol} describes how the non-Markovian dynamics of the system depends on the microscopic motion of the IO bath constituents.

The dynamics of $x(t)$ is determined by the function $F_N(t)$ defined in Eq.~\eqref{FN}. $F_N(t)$ is stochastic since it depends on the initial conditions $q_k(0)$ of the environmental oscillators which are unknown. We then see that, as expected, $x(t)$ displays the same diffusive behavior as the one given by stochastic force $g(t)$ in Eq.~\eqref{GLE}. Moreover, $F_N(t)$ displays two different contributions: the one given by $f_N(t)$ describes the behavior of the effective modes as if they where free oscillators. The second contribution depends on the interaction among the modes and the dynamics of the full chain. An important feature of Eq.~\eqref{xsol} is the way the peculiar environmental structure and the non-Markovian features show up: on the one side each $X_i$ contributes to the dynamics via the integral kernel $K_{i+1}$, reflecting the physical ordering among the effective modes. On the other side, the kernels themselves have a nested structure [see Eq.~\eqref{Ki}]. It is indeed this particular structure that paves the way for obtaining an effective description of the non-Markovian dynamics. One should also mention that such a structure is similar to the one obtained by Mori with his continued fraction description of the generalized Brownian motion~\cite{Mor65,Lee82}. This suggests that our derivation is a particular (and exactly solvable) case of the general projective technique proposed by Mori. However, we could not provide any direct evidence for this.



As previously mentioned, the GLE~\eqref{GLE} is the reference result for non-Markovian dynamics. It is then essential to compare our result to the GLE and show which are the advantages given by the microscopic description. 
One major difference is that the GLE is an integro-differential equation that needs to be solved, and remarkably the solution method (and its succesfullness) strongly depends on the spectral properties of the bath. In other words, the solution cannot be easily obtained for every set of environmental parameters. On the contrary, Eq.~\eqref{xsol} is the solution of Eq.~\eqref{xNint}, and it explicitly displays the dynamics of the system for any environment. 
Furthermore, since by construction the dynamics of the IO and chain models are equivalent, Eq.~\eqref{xsol} is also the solution of the GLE~\eqref{GLE} for an harmonic oscillator. 
This result gives an important contribution to the field as it allows for the exact treatment of a wider class of environments. Moreover, besides extending the range of applicability of the GLE, Eq.~\eqref{xsol} has also the value of being easier to analyze both at the analytical and numerical levels. 
It is worth mentioning that while in Eq.~\eqref{GLE} one can quite easily perform the Markov limit, obtaining the Langevin equation in the classical regime, one cannot do this as easily in Eq.~\eqref{xsol}. The GLE and the chain equation hence are two complementary descriptions, from which different effective descriptions can arise: the first is suitable to understand how the environment phenomenologically affects the system, the second to understand how non-Markovian features emerge from the underlying motion on short timescales. 
How such an effective description of non-Markovian dynamics can be realized will be analyzed in detail in a subsequent paper.

\section{Conclusions}

In order to provide a microscopic description of non-Markovian dynamics, we transformed the IO environment in a chain fashion, by solving an inverse eigenvalue problem. We derived the exact equation of motion for an harmonic oscillator, in terms of the microscopic motion of the environmental modes. This result represents a fundamental step forward in the understanding of non-Markovian dynamics. Equation~\eqref{xsol} shows how the microscopic behavior of the bath influences the system: the environmental oscillators act in a specific collective way, defining some collective modes. An interesting feature of the dynamics is that each collective mode $X_i$ acts through a kernel $K_{i+1}$: this peculiar structure will be subject of further studies.
Remarkably, Eq.~\eqref{xsol} provides the solution of the most important phenomenological equation describing general open quantum systems. This result allows to shed new light on the understanding of non-Markovian phenomena, allowing for an easier and deeper analysis both at the numerical and analytical levels.


\section{Acknowledgements}
The work of LF was supported by the Marie Curie Fellowship PIEF-GA-2012-328600. The authors wish to thank A. Bassi for many valuable discussions. LF further acknowledges A. Bassi for his hospitality at the University of Trieste.

\end{document}